\documentclass{ws-ijmpd}

\begin{document}

\markboth{M. de Avellar and J. E. Horvath} {Exact and quasi-exact
models of strange stars}

%
\catchline{}{}{}{}{}
%

\title{EXACT AND QUASI-EXACT MODELS OF STRANGE STARS}

\author{MARCIO G. B. DE AVELLAR$^{\star}$}
\author{J. E. HORVATH$^{\diamond}$}

\address{Instituto de Astronomia, Geof\'\i sica e Ci\^encias Amosf\'ericas, Universidade de S\~ao Paulo, \\
S\~ao Paulo 05508-090/SP,
Brazil\\
$\star$marcavel@astro.iag.usp.br\\$\diamond$foton@astro.iag.usp.br}

\maketitle

\begin{history}
\received{Day Month Year}
\revised{Day Month Year}
\comby{Managing Editor}
\end{history}

\begin{abstract}
We construct and compare a variety of simple models for strange
stars, namely, hypothetical self-bound objects made of a cold
stable version of the quark-gluon plasma. Exact, quasi-exact and
numerical models are examined to find the most economical
description for these objects. A simple and successful
parametrization of them is given in terms of the central density,
and many differences among the models are explicitly shown and
discussed.
\end{abstract}


\section{Introduction}

We study in this paper a variety of models for strange stars,
namely compact objects constituted of an hypothetic cold version
of a quark plasma featuring (by hypothesis) a lower energy per
baryon than normal hadronic matter. Quarks were suspected to
constitute a substantial fraction of a ``neutron'' star interior
since the early '70s\cite{itoh}\cdash\cite{collinsPerry}, and
their main properties explored. In 1984\cite{witten} the
possibility of a star fully composed of quarks up to its surface
(but perhaps covered with a ``normal'' crust) was considered. The
simplest model was constructed with a free quark+vacuum energy
equation of state (see Ref. \refcite{jaffe}). Because the scale
of the strong interactions is set by a multiple of the 
parameter $B$ of the MIT Bag Model, identified as the vacuum energy
density and numerically valued around $60 MeV \, fm^{-3}$ or
so\cite{fridolin_weber_livro}, these compact objects feature masses
$\sim 1M_{\odot}$ and radii $\sim 10 km$, pretty similar to their
``normal'' cousins. In fact they may constitute all the
``neutron'' stars hitherto observed.

The strong interactions (not gravity) bind these kind of compact objects,
and at least for the low-mass range $M \leq \, 1 \, M_{\odot}$, are the
responsible for the striking behavior of their mass-radius plot,
namely the continuity of models all the way down to
pretty small masses $\sim 0.5M_{\odot}$, the minimum allowed by the MIT equation of state.
Even though the low-mass stars are clearly
newtonian, gravity is increasingly strong at the higher end and
forces the use of general relativistic models when
$\frac{2GM}{c^{2}R} \geq 0.2 $.

As is well-known, the models of these objects can be described, in
a first approximation, by the solutions of the Einstein equations
for a perfect isotropic, static and spherically symmetric fluid
(whose metric is given by
$ds^{2}=c^{2}e^{\nu(r)}dt^{2}-e^{\lambda(r)}dr^{2}-r^2d\theta^{2}-r^{2}sin^{2}(\theta)d\phi^{2}$):

\begin{equation}
\frac{8\pi G}{c^2}\rho(r)=\frac{\lambda 'e^{-\lambda}}{r}+\frac{1-e^{-\lambda(r)}}{r^2},
\label{eq_eins_1}
\end{equation}

\begin{equation}
\frac{8\pi G}{c^4}p(r)=\frac{\nu 'e^{-\lambda}}{r}-\frac{1-e^{-\lambda(r)}}{r^2},
\label{eq_eins_2}
\end{equation}

\begin{equation}
p'(r)+\frac{1}{2}(c^{2}\rho(r)+p(r))\nu '(r)=0,
\label{eq_eins_3}
\end{equation}

where the last equation (contracted Bianchi identity) expresses
the conservation of energy-momentum of the fluid.

Mathematically, there are at least three different strategies to
solve these equations

\begin{itemize}
\item if the pressure, or the density or one metric element is
given (for example, by an {\it ansatz}), an exact or numerical
solution can be found by integration. However, this does not
guarantee to have any control over the equation of state
$p(\rho)$, and similar solution functions

\item instead, if the equation of state $p(\rho)$ is given (i.e.
the fluid is characterized from the beginning), the integration can
be performed (at least numerically) and the properties of the
stellar models follow

\item if both the equation of state and one additional function
($p$, $\rho$ or one of the metric) are known, a match
of the overdetermined system can be attempted, being in general
possible for some values of the involved parameters only.

\end{itemize}

It is not difficult to envision that the third route can be also
employed {\it without} an overdetermined system since one provide more
degrees of freedom (for example, electric field or
pressure anisotropy) at the expense of modifying the originally
posed physical problem. We shall return to this point below.


In this work, we have performed three sort of studies:

\begin{itemize}
\item the modeling of a strange star by using some of the well known exact solutions
(only that ones that meet physical requirements) for the Einstein Field Equation for
static and spherically symmetric perfect fluid, e.g., the solutions obtained by the first strategy;
\item the modeling of a strange star by the third routine, e.g. overdetermining the system,
in order to not only have the microphysics involved (by means of the equation of state),
but also to have control over the density profile, chosen to be gaussian as a first reasonable guess;
\item a comparison of the above studies to the true solution, or profiles, for a strange
star. By true solution, we mean that one obtained closing the system with the equation of state
of MIT Bag Model (see below) and numerically integrating the system.
\end{itemize}

With this we show that the applicability of the well known analytical solutions
are rather limited for modeling these stars and the best one can do having in mind
the original system is to construct a analytical quasi-exact solution with a overdetermined
system, by the expense to introduce an error in the geometry.

Then we show how the introduction of more degrees of freedom (specially pressure anisotropy
and the existence of an electric field) can avoid an overdetermined
system and at the same provide self-consistent analytical solutions for strange stars.

\section{Calculations}

\subsection{General considerations}

\begin{figure}[h!]
 \centering
 \includegraphics[scale=0.525]{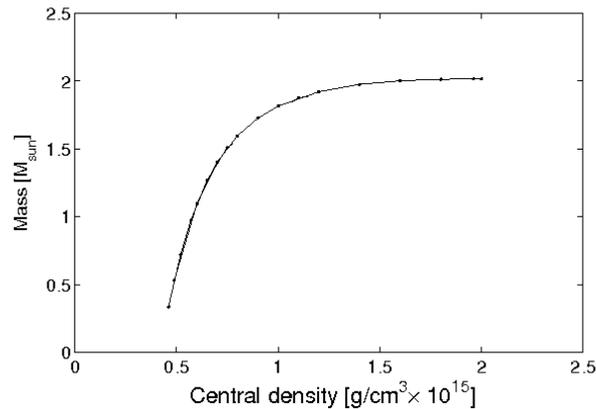}
 \caption{Mass-central density relation of the strange star sequence
 calculated by Alcock, Farhi and Olinto Ref. 6}
 \label{M_rhoc_AFO}
\end{figure}

\begin{figure}[h!]
 \centering
 \includegraphics[scale=0.525]{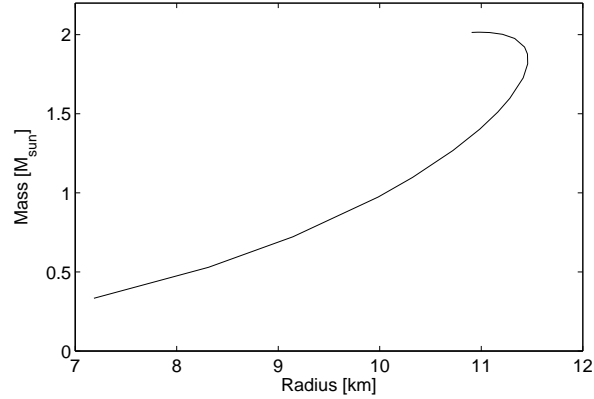}
 \caption{Mass-radius relation of the strange star sequence
 calculated by Alcock, Farhi and Olinto Ref. 6}
 \label{M_R_AFO}
\end{figure}

\begin{figure}[h!]
 \centering
 \includegraphics[scale=0.525]{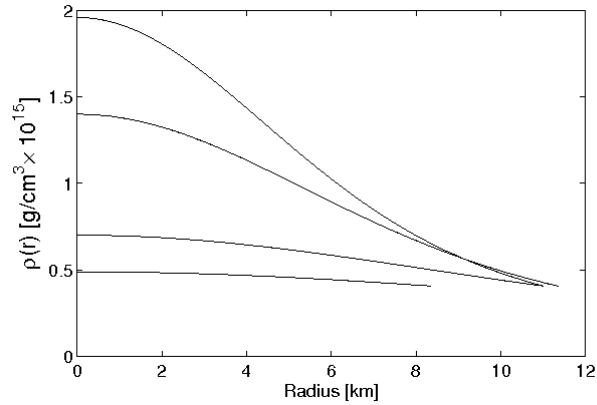}
 \caption{Density profiles of four selected strange stars
 calculated by Alcock, Farhi and Olinto Ref. 6}
 \label{perfil_dens_AFO}
\end{figure}

As a benchmark for our calculations we have adopted the simplest
numerical integration of the massless non-interacting quarks
presented, for example, by Alcock, Fahri and Olinto (hereafter
AFO) (Ref. \refcite{afo}). This is mainly motivated by the shapes
of the profiles obtained by them, which are reproduced in figures
\ref{M_rhoc_AFO}, \ref{M_R_AFO} and \ref{perfil_dens_AFO}. It is
clear that any other model in which interactions or finite-quark
masses are needed can be obtained analogously (see Ref.
\refcite{jorge} for an exploration of the full parameter space of
the models).

Since the main goal of this work is to simplify the description of
the stellar models as much as possible, possibly neglecting small
($\leq 1\%$) differences arising from the fact that numerical and
exact (or quasi-exact) models do not match exactly, it is useful
to parametrize all the features in terms of a single input
parameter, chosen to be the central density $\rho_{c}$ which must
satisfy $\rho_{c} >4 B/c^{2}$ in order to be consistent with the
description of the matter in the {\it MIT bag model}, where
$p=(c^{2}\rho-4B)/3$.

To fix ideas, assume that $f(x)=a g(x)+b$ is
a solution of the density profile. If an {\it ansatz} is given for $g(x)$, 
with that profile it is possible to obtain
solutions for the pressure and the for the metric elements. Then we fit the parameters
$a=a(\rho_{c})$ and $b=b(\rho_{c})$, so that the mass-radius
relation of AFO is reproduced as accurately as possible. After this is
achieved, we can proceed to derive all the features of any given
model.

However, not all exact (or even quasi-exact, see below) solutions
describe a viable star: physically acceptable models should
fulfil the following conditions, as discussed in the work by
Delgaty and Lake (Ref. \refcite{catalogo_delgaty_lake}):

\begin{itemize}
\item $e^{\lambda(0)}=1$

To integrate the Einstein equations, two initial conditions are
needed, generally taken as $m(r=0)$ and $p(r=0)$, this is why this
condition is so important to define the central region of the
star. A small sphere of radius $r=\epsilon$ has a length of
$2\pi\epsilon$ and proper radius $e^{\lambda /2}\epsilon$. A small
circle around $r=0$ has a ratio of the length to its radius given
by $2\pi e^{-\lambda /2}$. However, since the space-time is
locally flat, this ratio should be for this small circle around
$r=0$, equal to $2\pi$. Thus, $e^{\lambda(0)}=1$ and if
$r\rightarrow 0$, then also $m(r)\rightarrow 0$, cf. Ref.
\refcite{schutz};

\item $e^{\nu(0)}=constant<1$

A particle characterized by the geodesic with a given constant
$\mathcal{E}$ has an energy $e^{-\nu /2}\mathcal{E}$ relative to a
locally inertial observer at rest in this space-time. Since
$e^{-\nu /2}=1$ far away from the object, $\mathcal{E}$ is the
energy a distant observer would measure if the particle was
distant (i.e. the energy at infinity). Since $e^{-\nu /2} > 1$ for
all other regions not at the infinite\footnote{From the
continuity condition at the boundary of the object,
$e^{\nu(r=R)}=1-\frac{2GM}{c^{2}R}<1$. But
$e^{\nu(r>R)}=1-\frac{2GM}{c^{2}r}\rightarrow 1$ if $r\rightarrow
\infty$.}, the observer will measure a higher energy than
$\mathcal{E}$ in that regions. This extra energy is the energy a
particle gains when it falls in the gravitational field of the
object. Thus, $e^{-\nu /2}>1\Rightarrow e^{\nu(0) /2}<1$, cf. Ref.
\refcite{schutz};

\item $\rho(r)$ regular at the origin and positive definite

A region around $r=0$, small enough to be called ``center of the
star'' and large enough to maintain the physical features of the
fluid at constant (uniform) density, must have a positive value for
the fluid to be real;

\item $p(r)$ regular at the origin and positive definite

The pressure at a central core which is approximated by a constant
(uniform) density must have a constant positive value, as a
result of the properties of real matter;

\item $\frac{dp}{d\rho}\leq c^{2}$

The speed of sound in the fluid should be smaller than the velocity of light.
\end{itemize}

Our first approach to the problem is described below, using the
well-known Tolman IV\cite{tolman_1939} and Buchdhal I\cite{buchI},
which satisfy all properties 1-5 necessary for realistic models.

\subsection{Exact and quasi-exact solutions}

\subsubsection{Tolman IV and Buchdahl I}

Tolman IV and Buchdahl I are two well-known exact solutions
featuring mathematically simple expressions for static
self-gravitating fluid spheres. Moreover, they are physically sound 
and fulfil tghe requirements listed in the former section. 
We have tested the applicability
of these simple models to the strange star problem, comparing them
with the numerical results of AFO. A previous attempt can be found
in Ref. \refcite{Latt}.\\

We chose these two solutions also because the density profile can be
well fitted in order to reproduce the AFO density profile.

\indent {\bf Tolman IV}

This solution has been found by R. Tolman in his seminal paper (\refcite{tolman_1939}).
To solve the Einstein equations, he made an {\it ansatz}
$e^{\nu}\nu^{'}/2r=const$, a method that rendered an exact
integration of the problem. The functions thus obtained read

$$
e^{\lambda}=\frac{1+2r^{2}/A^{2}}{(1-r^{2}/R^{2})(1+r^{2}/A^{2})},
$$

$$
e^{\nu}=Q^{2}\Big(1+\frac{r^{2}}{A^{2}}\Big),
$$

$$
\frac{8\pi G}{c^{2}}\rho(r)=\frac{1}{A^{2}}\frac{1+3A^{2}/R^{2}+3r^{2}/R^{2}}
{1+2r^{2}/A^{2}}+\frac{2}{A^{2}}\frac{1-r^{2}/R^{2}}{(1+2r^{2}/A^{2})^{2}}
$$
and

$$
\frac{8\pi G}{c^{4}}p(r)=\frac{1}{A^{2}}\frac{1-A^{2}/R^{2}-3r^{2}/R^{2}}{1+2r^{2}/A^{2}}.
$$

From these expressions, an equation of state, a physical radius
(boundary, $r_{b}$) the constant {\it Q} and the total mass of the sphere
are, respectively

$$
\rho(p)=\rho_{c}-5\frac{p_{c}}{c^{2}}+5\frac{p}{c^{2}}+8\frac{(\frac{p_{c}}{c^{2}}-
\frac{p}{c^{2}})^{2}}{\frac{p_{c}}{c^{2}}+\rho_{c}},
$$

$$
r_{b}=\frac{R}{3^{1/2}}\Big(1-\frac{A^{2}}{R^{2}}\Big)^{1/2},
$$

$$
Q^{2}=\frac{1-r_{b}^{2}/R^{2}}{1+2r_{b}^{2}/A^{2}},
$$

$$
M_{bound.cond.}=\frac{c^{2}r_{b}}{2G}\Big(1-\frac{(1-r_{b}^{2}/R^{2})
(1+r_{b}^{2}/A^{2})}{1+2r_{b}^{2}/A^{2}}\Big),
$$ with

$$
\frac{8\pi G}{c^{2}}\rho_{c}=3\frac{R^{2}+A^{2}}{R^{2}A^{2}},\hspace{1.0cm}
\frac{8\pi G}{c^{4}}p_{c}=\frac{R^{2}-A^{2}}{A^{2}R^{2}}
$$ with $A$, $R$ and $Q$ three arbitrary constants.

\begin{figure}[h!]
 \centering
 \includegraphics[scale=0.625]{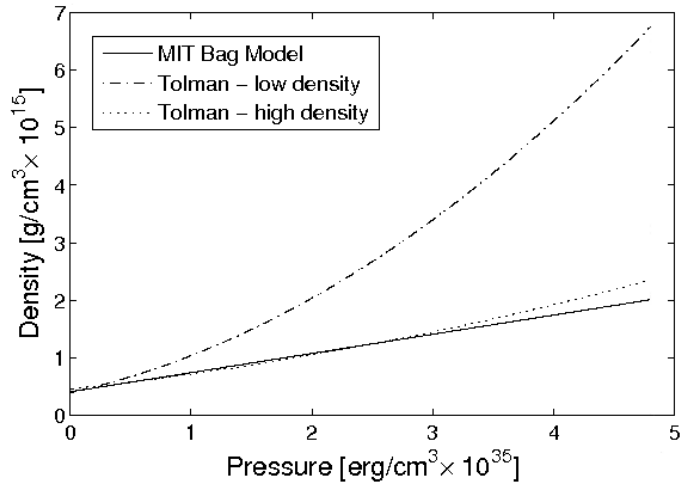}
 \caption{Variation of the equation of state with the constants $A , R$ present in Tolman's models Ref. 10}
 \label{ilustrando_EOS}
\end{figure}

To match numerical strange star models as calculated by AFO with
the Tolman IV solution, we seek to reproduce the mass and the
radius of each model, that is, to determine $A$, $R$ and $Q$ so
that the mass and the radius of the numerical calculations are
reproduced. However, the central density could not be adjusted
simultaneously (the later depends on $A$ and $R$, which in turn are
determined by the radius and mass). A relevant question is whether
the resulting equation of state is compatible with the linear
expression of the {\it MIT Bag Model}. We can easily check that
the Tolman IV is not appropriate, first because it is not linear,
but more importantly because that it depends on the $A$ and $R$
present in the metric elements. When $A$ and $R$ are adjusted to
reproduce accurately the masses and radii of the AFO calculations,
they happen to depend on the central density of the model.
Therefore, the equation of state varies from model to model (from star to star), in
spite that its functional form remains the same. This behavior is
illustrated in Fig.\ref{ilustrando_EOS}.

It is clear that an equation of state that varies from model to
model is not useful to describe a compact star, since the equation
of state should reflect the fundamental properties of cold matter
and should not behave in this way.\\

\indent {\bf Buchdahl I}

Much in the same way as in Tolman IV, we examined the Buchdahl I\cite{buchI}
solution. The expressions are given by

$$
e^{\lambda}=\frac{2(1+Cr^{2})}{2-Cr^{2}},
$$

$$
e^{\nu}=A\big[(1+Cr^{2})^{3/2}+B\sqrt{2-Cr^{2}}(5+2Cr^{2})\big]^{2},
$$

$$
\frac{8\pi G}{c^{2}}\rho(r)=-\frac{-\frac{2Cr}{2+2Cr^{2}}-\frac{4Cr(2-Cr^2)}
{(2+2Cr^{2})^{2}}}{r}+\frac{1-\frac{2-Cr^2}{2+2Cr^{2}}}{r^2},
$$

$$
\frac{8\pi G}{c^{4}}p(r)=\frac{2\big[3Cr\sqrt{1+Cr^{2}}-\frac{BCr(5+2Cr^{2})}
{\sqrt{2-Cr^{2}}}+4BCr\sqrt{2-Cr^{2}}\big](2-Cr^2)}{[(1+Cr^{2})^{3/2}+B\sqrt{2-Cr^{2}}
(5+2Cr^{2})](2+2Cr^{2})r}
$$

$$
\hspace{4.5cm} -\frac{1-\frac{2-Cr^2}{2+2Cr^{2}}}{r^2},
$$ where $A$, $B$ and $C$ are arbitrary constants.

Two types of fits were possible: in the first one we fitted the
mass and radius of each model, and as a consequence $\rho_{c}\neq
\rho_{c_{AFO}}$ as before. We also attempted to fit both the mass
and central density, leading to different radii, depending on the
masses. However, we eventually came to the same problem of the
non-linearity of the equation of state. Again the obtained
equation of state depends on variable quantities and is not useful
to model the quark matter.

\subsubsection{Quasi-exact gaussian model}

An inspection to the density profiles presented by AFO (Figure
\ref{perfil_dens_AFO}) suggests a simple and popular
parametrization of the density dependence. As a simple {\it
ansatz} we assumed a gaussian profile, a procedure that led to an
overdetermination of the system (since the linear MIT equation of
state was also imposed). With those assumptions, the following
expressions were immediately obtained in closed forms

\begin{equation}
\rho(r)=\Big(\rho_{c}-\frac{B}{c^2}\Big)e^{-r^{2}/r_{o}^{2}}+\frac{B}{c^2},
\end{equation}

\begin{equation}
p(r)=c^{2}\rho(r)-\frac{4B}{3}
\end{equation}

\begin{equation}
\nu(r)=\frac{r^2}{2r_o^2}+a.
\end{equation}

$$
e^{-\lambda(r)}=\frac{4\pi G\rho_{c}e^{-r^{2}/r_{o}^{2}}r^{2}_{o}}{c^{2}}-
\frac{2\pi^{3/2}G\rho_{c}r^{3}_{o}erf(r/r_{o})}{rc^{2}}-\frac{4\pi GBe^{-r^{2}/r_{o}^{2}}r^{2}_{o}}
{c^{4}}
$$

\begin{equation}
\hspace{2.5cm} +\frac{2\pi^{3/2}GBr^{3}_{o}erf(r/r_{o})}{rc^{4}}-\frac{8\pi Gr^{2}B}{3c^{4}}+1,
\end{equation}

\begin{equation}
e^{-\lambda(r)}=\frac{r^{2}_{o}\Big(8\pi Gr^{2}e^{-r^{2}/r^{2}_{o}}
\rho_{c}c^{2}-8\pi Gr^{2}e^{-r^{2}/r^{2}_{o}}B-24\pi Gr^{2}B+3c^{4}\Big)}{3c^{4}(r^{2}+r^{2}_{o})}.
\end{equation}

The problem posed in this way reduces to prove whether the two
expressions for $\exp(\lambda(r))$, eqs. (7) and (9), are equivalent. Alternatively,
even if the functions are different, they could be almost
identical inside the star, and therefore a small error would be
introduced by employing either one. This is why we speak about
quasi-exact solutions for the linear equation of state to stellar problem.

We used as benchmark the same four density profiles shown in AFO
for definite values of the total mass $M=1.99$ $M_{\odot}$,
$M=1.95$ $M_{\odot}$, $M=1.4$ $M_{\odot}$ and $M=0.53$
$M_{\odot}$.

The idea was to fit expressions for $a$ and for $r_o$ (the scale of decay of the density profile) as
a function of the central density $\rho_{c}$ which could reproduce
the masses and radii.

After that, given any $\rho_{c}$, one may
obtain any stellar model along the sequence because he can
calculate the total mass, the radius, the metric elements and the
density and pressure profiles.

However, it proved impossible to fit all parameters simultaneously
without making the system inconsistent. Letting the radius to vary
(i.e. relaxing the condition for them to reproduce the AFO
results), but keeping the masses and central densities, then the stellar
radius $R$, the parameter $a$ and the scale $r_o$ could be obtained.
As an important remark, we stress that fixing the mass means that
the integral of the density function and the boundary condition
derived from the metric $\nu(r)$ must be the same, with their
values the same as the ones found by AFO.

The fit resulting from just the four profiles did not allow a good
determination of $a$ and $r_{o}$. Thus, the grid of models was
extended to 20 profiles (i.e. 20 values of the central density) to
cover wider mass and radii intervals. The values of $r_o$ and $a$
are shown in Table\ref{ro_a}.

\begin{table}
\begin{center}
\begin{tabular}{|c|c|c|c|}
\hline
$\rho_c [10^{14}g/cm^3]$ & $M/M_{\odot}[AFO]$  & $r_o [km]$ & $a$ \\
\hline
21 & 2.014042750 & 7.631010387 & -1.780521357 \\
20.5 & 2.014801608 & 7.714337813 & -1.762999185 \\
20 & 2.015326779 & 7.801045927 & -1.744982347 \\
19.6 & 2.015436678 & 7.872850429 & -1.730140677 \\
18 & 2.013326957 & 8.185473708 & -1.666867564 \\
16 & 2.002284428 & 8.646108988 & -1.577075908 \\
14 & 1.975792515 & 9.212929519 & -1.47142208 \\
12 & 1.921914333 & 9.936363686 & -1.343215145 \\
11 & 1.877755062 & 10.38285995 & -1.267268177 \\
10 & 1.815572219 & 10.90882098 & -1.180584905 \\
9 & 1.726746594 & 11.54174779 & -1.079793495 \\
8 & 1.597234612 & 12.32531382 & -0.959831395 \\
7.5 & 1.510155564 & 12.79466558 & -0.890275006 \\
7 & 1.402444810 & 13.33296263 & -0.81247546 \\
6.5 & 1.267766569 & 13.95933096 & -0.724422383 \\
6 & 1.097507111 & 14.70148514 & -0.62334087 \\
5.7 & 0.973622557 & 15.21887892 & -0.554849121 \\
5.2 & 0.721604642 & 16.24225101 & -0.423820856 \\
4.885 & 0.529204878 & 17.02459731 & -0.327483126 \\
4.6 & 0.333488500 & 17.85691977 & -0.228380811 \\
\hline
\end{tabular}
\end{center}
\caption{Values of $r_o$ and $a$ obtained by imposing the same AFO
mass value} \label{ro_a}
\end{table}

Having these values for $a$ and $r_{o}$, we proceeded to derive an
accurate functional fit of the form,

\begin{equation}
a(w)=-2.4+9.96e^{(-\frac{w}{0.19})}+2.426e^{(-\frac{w}{0.59})}+1.92e^{(-\frac{w}{3.45})}
\end{equation}

\begin{equation}
r_{o}(w)=5.56+22.46e^{(-\frac{w}{0.50})}+10.35e^{(-\frac{w}{2.42})}
+169.46e^{(-\frac{w}{0.17})}[km],
\end{equation}

where $w\equiv\frac{8\pi G}{c^2}l^2\rho_{c}$, $l\equiv 10$ $km$,
is a dimensionless quantity of the order unity that embodies
physical constants ($c$, $G$ etc.) and the lengthscale of the
object ($\sim 10 km$).

\begin{figure}[htbp!]
\centering
{
    \includegraphics[width=8.90cm]{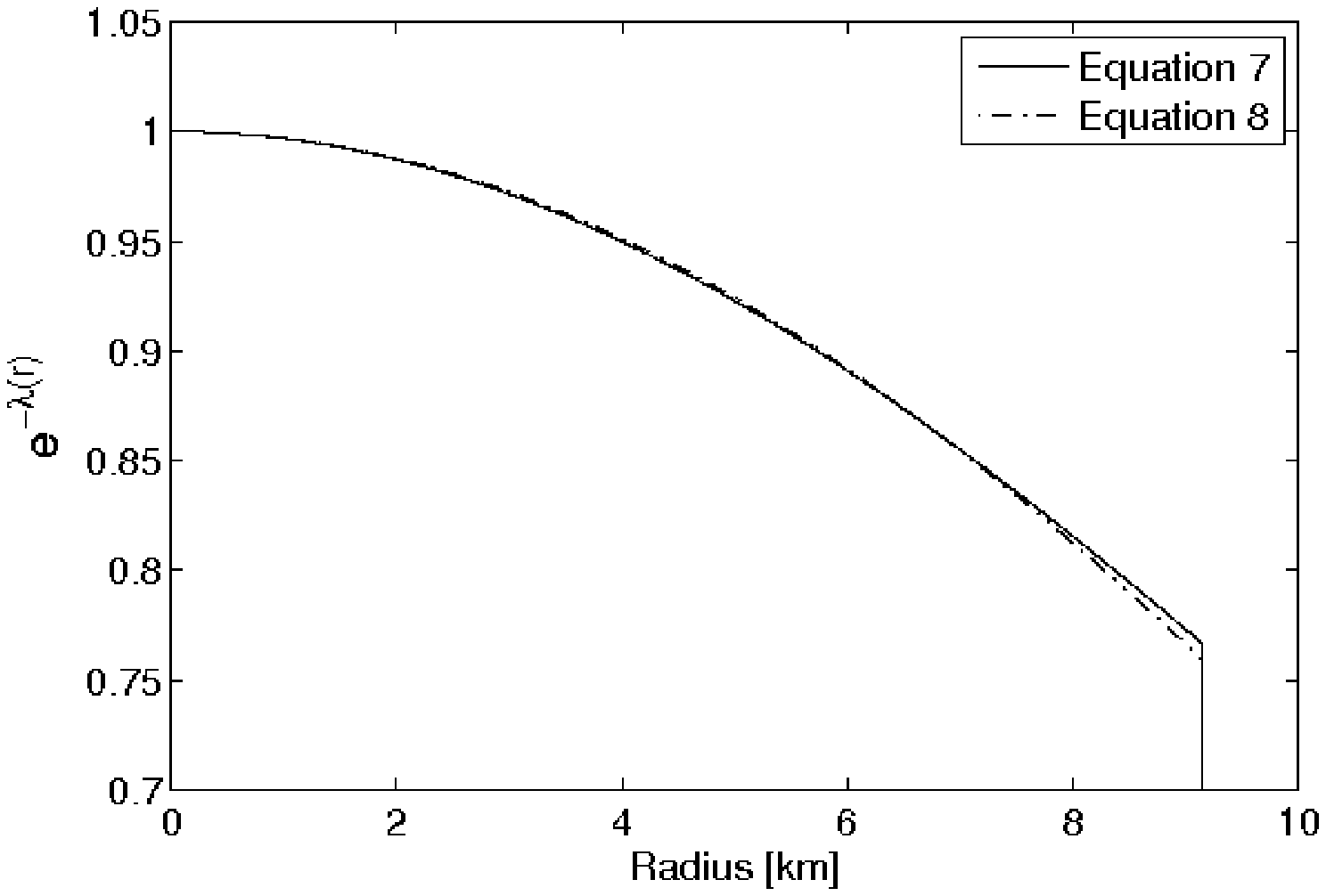}
} \hspace{1cm}
{
    \includegraphics[width=8.70cm]{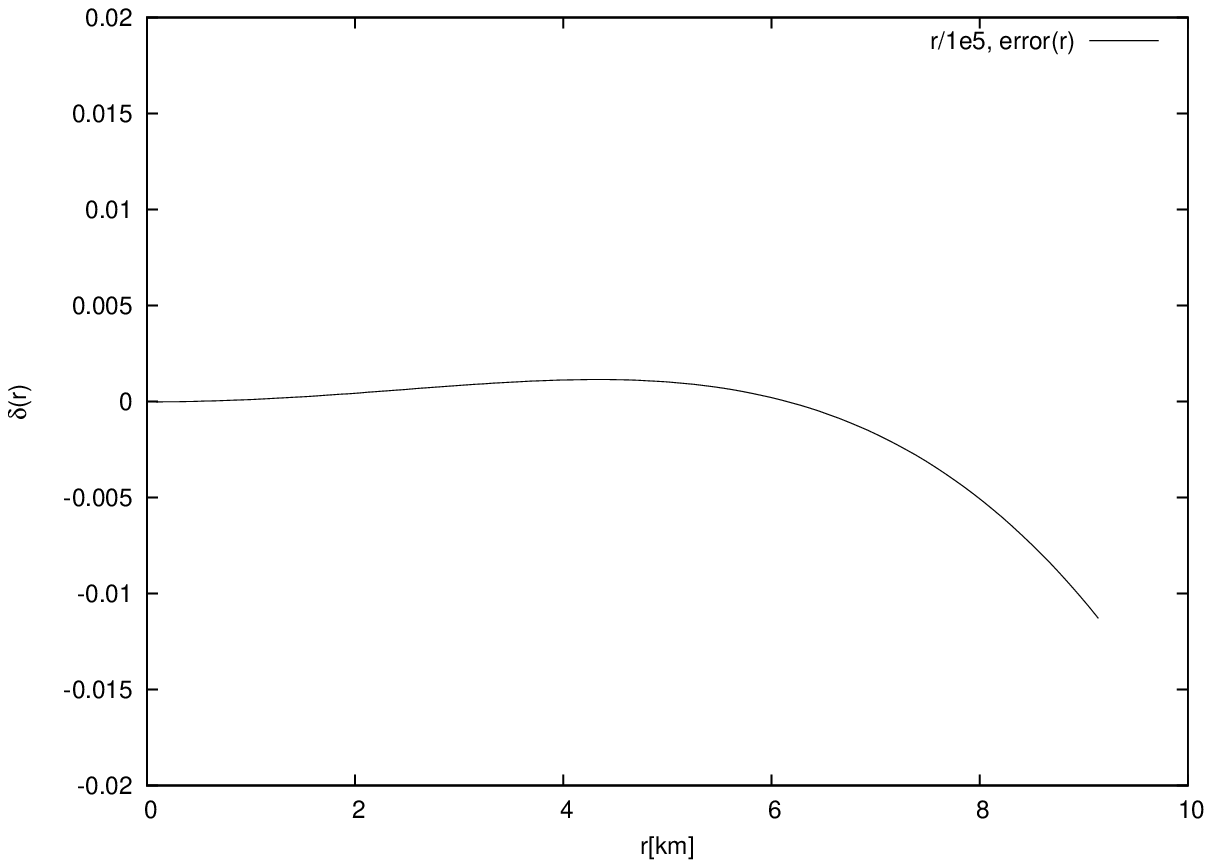}
} \caption{The difference between the two expressions of the metric element $e^{-\lambda(r)}$ inside the star for the model with $\rho_{c}=5.2\times10^{14}g/cm^{3}$.
The radius of the star is marked by the vertical line at the end of the curves.
Up: the two expressions; Bottom: the relative error as a function of the radial coordinate.} 
\label{baixa_dens}
\end{figure}

\begin{figure}[htbp!]
\centering
{
    \includegraphics[width=8.90cm]{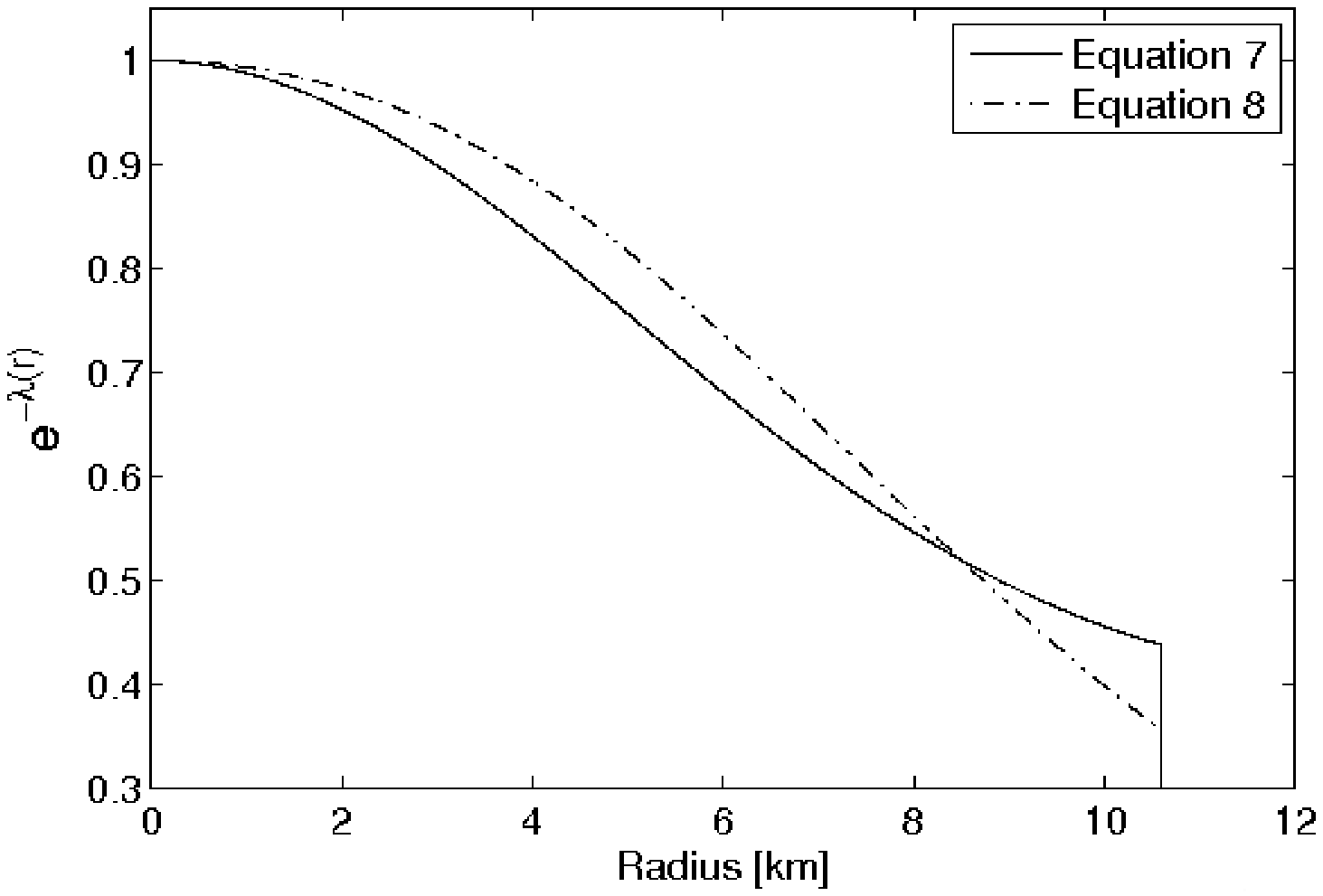}
} \hspace{1cm}
{
    \includegraphics[width=8.70cm]{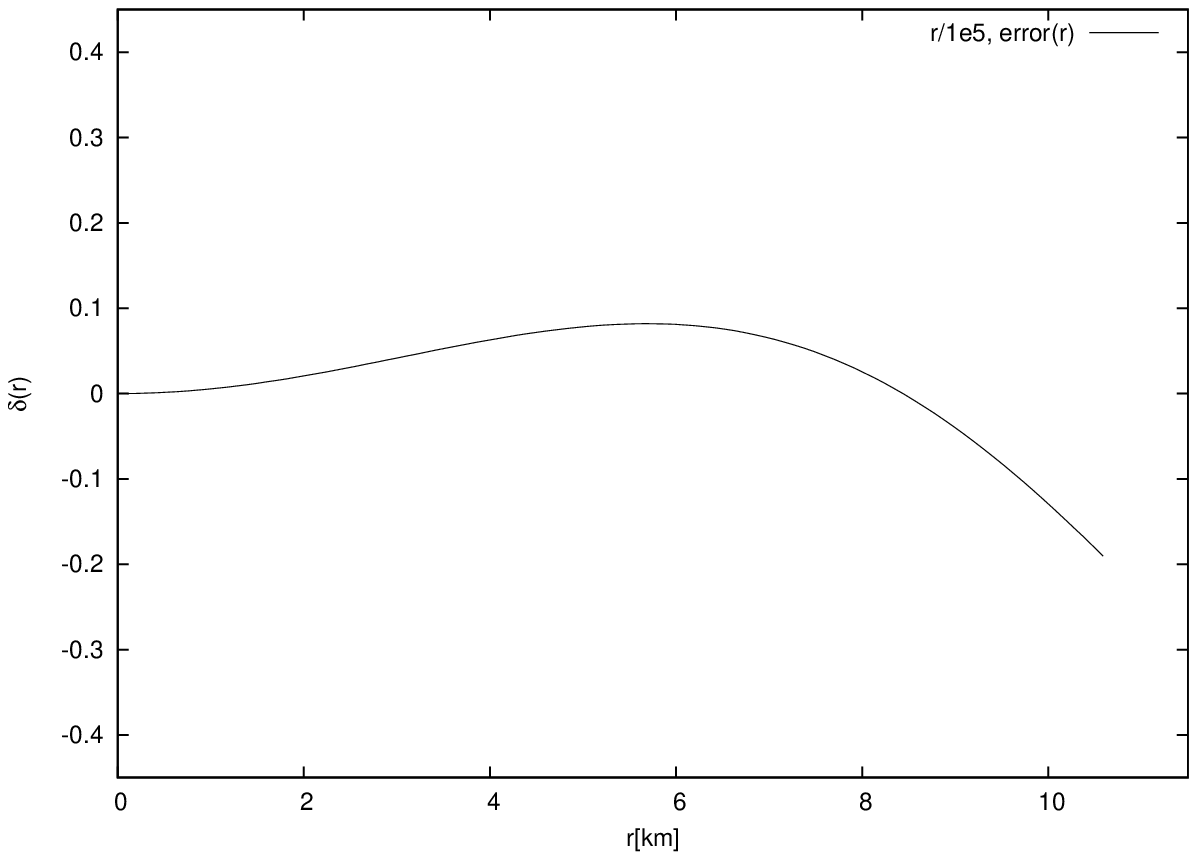}
} \caption{The same as in Fig. 5 for the model with
 $\rho_{c}=19.6\times10^{14}g/cm^{3}$. The differences are bigger,
 as pointed out in the text.} 
\label{alta_dens}
\end{figure}

With these values for $r_o$ and $a$, we may wonder about the
accuracy of the metric element $e^{\lambda(r)}$. We have seen that
an ambiguity for this quantity exists. Its first expression was
derived from the first Einstein equation (eq. \ref{eq_eins_1}) and
the second from the second one (eq. \ref{eq_eins_2}). How
different they really are? Actually it can be checked that the
agreement is very good for the lower densities (Figure
\ref{baixa_dens}), and worsens for the highest densities (Figure
\ref{alta_dens}) along the stellar sequence. This behavior is a
consequence of the newtonian character of the gravitational field
for the lower densities, already discussed in Section 1.

In summary, we have shown that with just the central density value
(quite analogously to any other calculation of stellar structure
that begins by specifying this value), the values of $a$ and
$r_{o}$ can be calculated from the analytical fit. They determine
in turn the profiles of density and pressure within the gaussian
{\it ansatz}, and with them the physical radius of the model $R$.
Finally, the mass can be calculated analytically by integrating
the gaussian density profile and/or using the boundary condition,
yielding the same values within $\lesssim 10\%$ because of the deviations
of the fitted expressions from the exact numerical results. We
believe that this quasi-analytical model (eqs. 4-8) can be useful for a
(very) accurate evaluation of the stellar model structure in a
variety of situations.

It is worth to mention at this point that Cheng and Harko\cite{cheng} presented a very
similar approach, starting with the relativistic conditions of thermodynamic
equilibrium and the MIT bag equation of state. They found very accurate approximate
mass and radius formula for strange stars, in the static case and, amazingly,
in the rotating case. Their errors are less than 1\% for the former case and 3\%
for maximally rotating stars.

Considering how the general relativity changes the conditions of thermal equilibrium inside the
compact star, they started with the equation for the chemical potential

\begin{equation}
\frac{d\mu}{\mu}=\frac{dp}{\epsilon+p}.
\end{equation}

Comparing the conditions at the center and at the boundary (surface)

\begin{equation}
\frac{\mu_{c}^{2}}{\mu_{s}^{2}}=\frac{1}{C(\rho_{c},B)}\Big(1-\frac{2GM}{c^{2}R}\Big)
\end{equation}

and with the MIT Bag equation of the state, they could integrate the equation 11 to obtain

\begin{equation}
\mu=\frac{\mu_{c}}{(\epsilon_{c}-B)^{1/4}}(\epsilon-B)^{1/4}.
\end{equation}

Here, the function $C$ is related to the time component of the metric tensor at the center of the star,
$g_{00}l_{c}=C(\rho_{c},B)$, that is also related to the redshift of a photon emitted at the center of
the star.

Thus, defining the dimensionless parameter $\eta\equiv\rho_{c}/B$, they obtained the following expression 
to the mass-radius ratio:

\begin{equation}
\frac{M}{R}=\frac{c^{2}}{2G}\Big[1-C(\eta)\Big(\frac{\eta-1}{3}\Big)^{1/2}\Big].
\end{equation}

From considerations about the mass continuity on the surface, they proposed the following mass-radius relation

\begin{equation}
M=\frac{16\pi B}{3}a(\eta)R^{3}.
\end{equation}

Here, $C(\eta)=const+\sum_{i}a_{i}/\eta^{i}$ ($a_{i}$ constants) and similarly to $a(\eta)$.

Finally, they integrated the Einstein Field Equations numerically and fitted $C(\eta)$ and $a(\eta)$ in order to
obtain expressions for the mass and radius in closed forms, where they used $B=56MeV/fm^{3}$.

Their formulae, however, become increasingly inaccurate when $\rho_{c}\rightarrow 4B$.

In the following, we give our derivation for mass and radius, the analogous to their formulae (see eqs. 14 and 15 
of Ref. 13),

\begin{equation}
R=\sqrt{-[r_{o}(w)]^2ln\Big(\frac{3B}{c^{2}\rho_{c}-B}\Big)}\hspace{0.2cm}[km]
\end{equation}

\begin{equation}
M=5\times10^{-29}\frac{\Big(1-e^{\big(\frac{1}{2}\frac{R^{2}}{[r_{o}(w)]^{2}}+a(w)\big)}\Big)c^{2}R}{2G}\hspace{0.2cm}[M_{\odot}]
\end{equation}

We stress two important points about these two expressions. First, they provide the ``right'' 
value for the zero-pressure situation, although this feature should not be taken too strictly, 
because the expressions are only approximations. Second, although they appear quite generic, 
one should remember that we used $B=57.5MeV/fm^{3}$ to derive $a(w)$ and $r_{o}(w)$. A different 
value of $B$ would force to recalculate the latter. 

The advantage of our approach is that we have the profile for $\rho(r)$, $p(r)$, $\nu(r)$ 
and $\lambda(r)$ in closed form for each possible value of the central density $\rho_{c}$.
Besides that, our formula are in complete agreement with the formula derived 
by Cheng and Harko\cite{cheng}.

We also point out that recently, Narain, Schaffner-Bielich and
Mishustin\cite{bielich} have made a numerical calculation
integrating the dimensionless TOV equation with a linear equation
of state for generic fermionic matter. The possibility of scaling the
solutions of the TOV problem was already known and exploited, for
example, in Witten's paper \cite{witten}. Using this property,
Narain, Schaffner-Bielich and Mishustin found a very useful
general scaling solution and discussed how to rescale these
equations in order to find the solutions for arbitrary fermion
masses and interacting strengths. In spite of this generality, we
point out that our process of making the equations dimensionless
is different from that work and, moreover, the choice of a
gaussian form induces the presence of the length $r_{o}$ which
controls the spatial decay of the density. As a result, it is not
possible to compare easily our quasi-exact models with their
results, which remain more general but require the knowledge of
the dimensionless $M-R$ curve.

\subsubsection{Polytropic models}

A recent interesting approach has been presented by Lai and Xu\cite{xu},
which models a quark star with a polytropic equation of state.
This kind of equation of state is generally stiffer than that
conventional linear one, like the bag model.  Unfortunately, exact
solutions are difficult to obtain for the polytropic case.
However, their numerical results are interesting because they deal
with two models: with and without QCD vacuum energy ($\Lambda$),
and show how they compare to linear bag-like equations of state.
See their numerical results in Figure 2 from Ref. \refcite{xu}.

As a general feature we may say that the work of Lai and Xu shows
that a polytropic description of strange stars is possible and
accurate, but it does not provide an easy and economical form for
generating models, at least not easier than a full numerical work.

\subsubsection{Exact anisotropic model}

The exact anisotropic model of Sharma and Maharaj\cite{sharma}
is another accurate approximation for the stellar structure and
needs, as above, just one parameter (the central density). As a
physical motivation to consider this anisotropy, they quote
Usov\cite{usov} that suppose a strong electric field formed into a
thin layer at the quark surface of a bare strange star. This is
possible due the depletion of s-quarks in this region.

\begin{figure}[htbp!]
\centering
{
    \includegraphics[width=8.75cm]{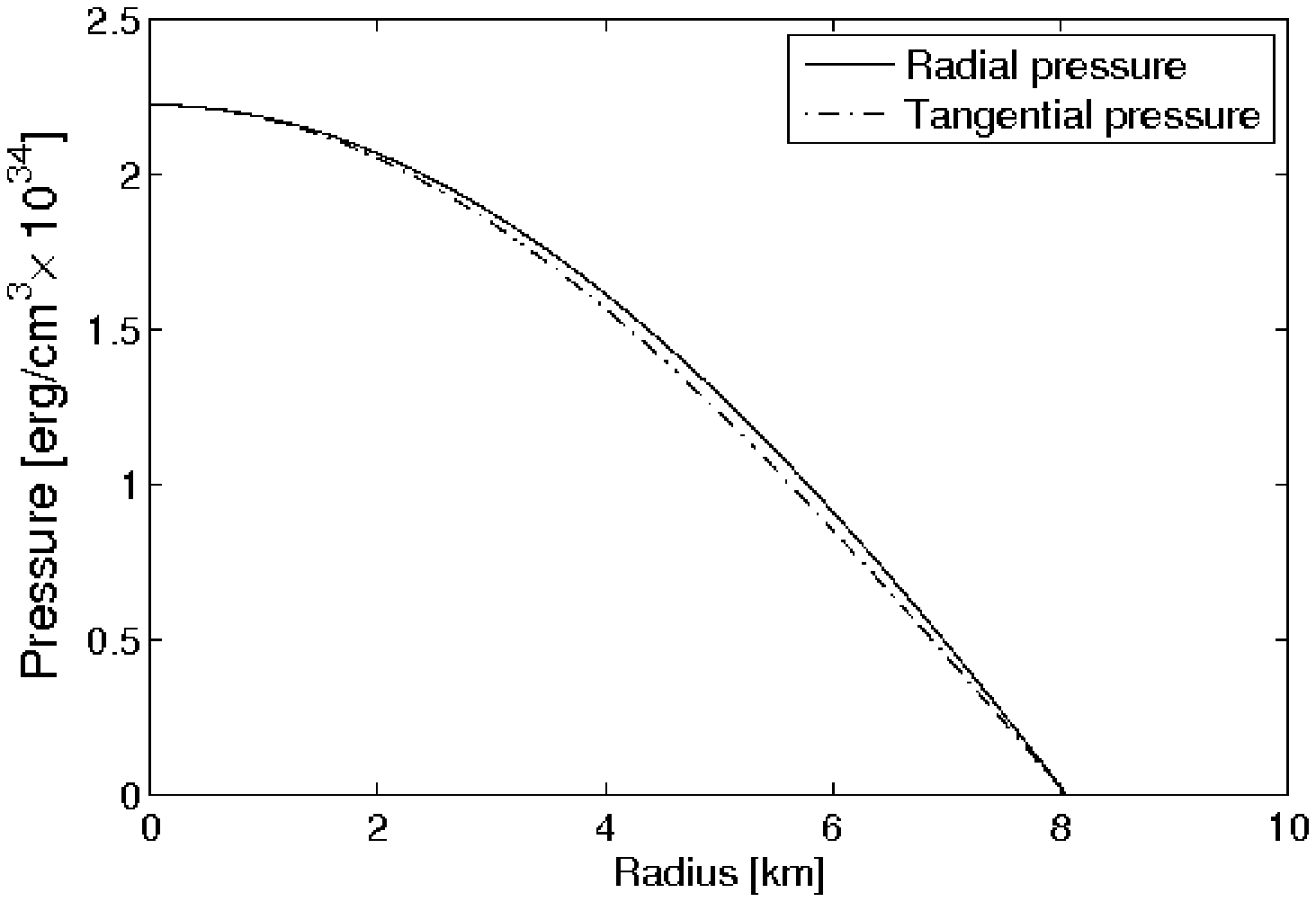}
} \hspace{1cm}
{
    \includegraphics[width=8.70cm]{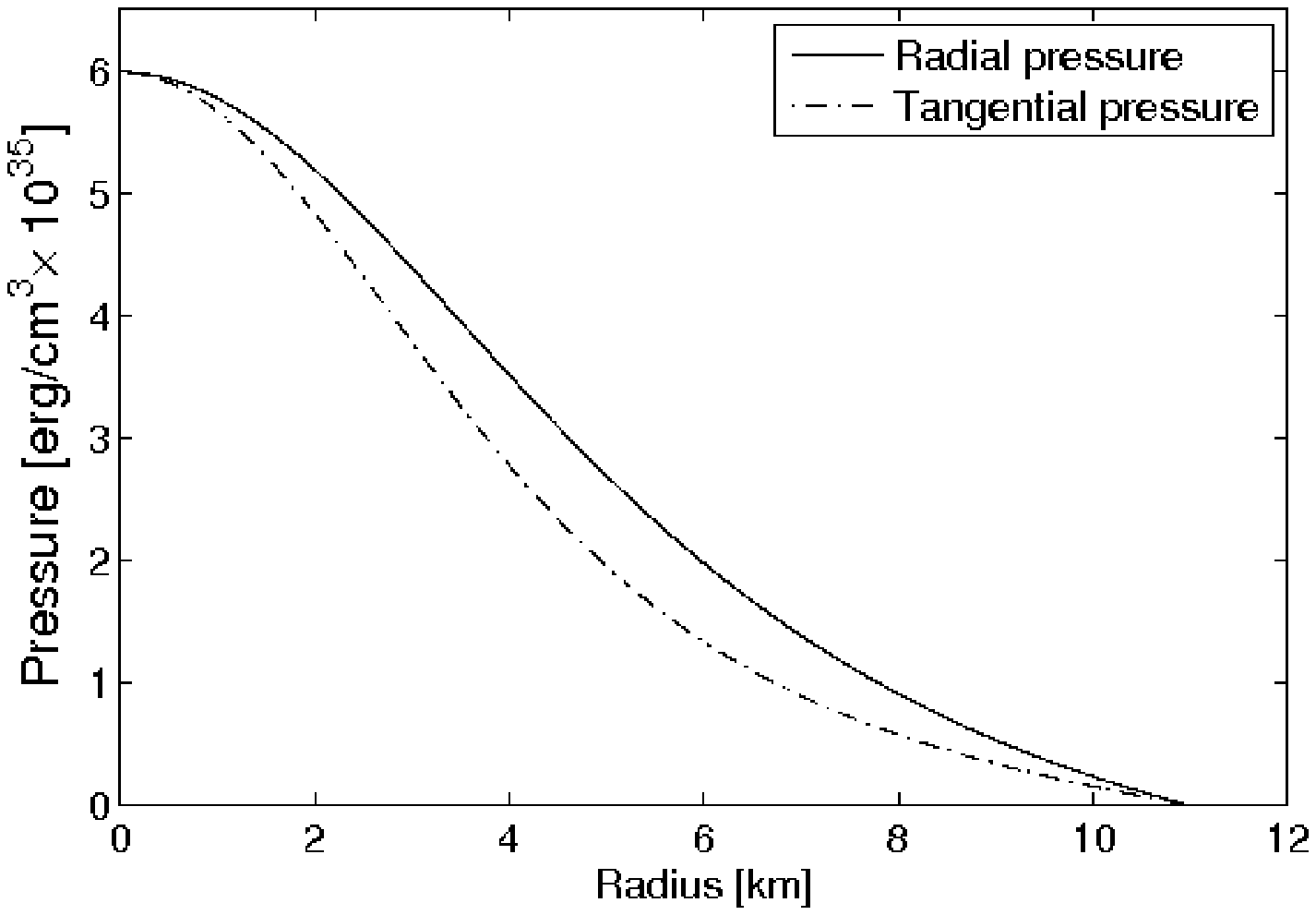}
} \caption{Radial and tangential pressures for two anisotropic stellar models.
Up: $\rho_{c}=4.8\times 10^{14}g/cm^{3}$; Bottom: $\rho_{c}=24\times 10^{14}g/cm^{3}$.} 
\label{pressoes}
\end{figure}

\begin{figure}[h!]
 \centering
 \includegraphics[scale=0.625]{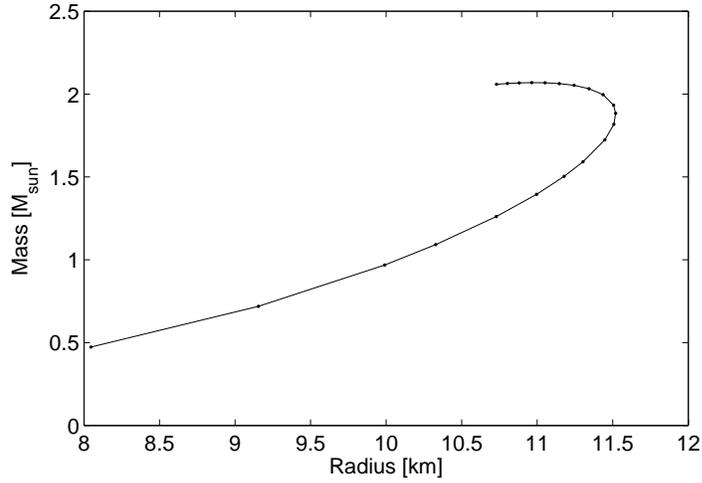}
 \caption{Mass-radius curve for the anisotropic model of Sharma-Maharaj Ref. 16}
 \label{massa-raio}
\end{figure}

The exact expressions of this models are

\begin{equation}
\rho(r)=\frac{b(3+ar^{2})}{8\pi (1+ar^{2})^{2}},
\end{equation}

\begin{equation}
m(r)=\frac{r^{3}b}{2(1+ar^{2})},
\end{equation}

\begin{equation}
p_{r}(r)=\frac{c^{2}b(3+ar^{2})}{24\pi(1+ar^{2})^{2}}-\frac{4B}{3},
\end{equation}

\begin{equation}
e^{\lambda(r)}=\frac{(1+ar^{2})c^{2}}{(c^{2}a-Gb)r^{2}+c^{2}},
\end{equation}

\begin{equation}
e^{\nu(r)}=e^{Xr^2}(1+ar^{2})^{1/3}(c^{2}+c^{2}ar^{2}-Gr^{2}b)^{Y}K
\end{equation}

$$
p_{t}(r)=p_{r}(r)-\frac{1}{2}\Bigg[-\frac{c^{2}bar}{12\pi
(1+ar^{2})^{2}}+ \frac{c^{2}bar(3+ar^{2})}{6\pi
(1+ar^{2})^{3}}
$$

\begin{equation}
\hspace{2.5cm} -\frac{(\frac{c^{2}b(3+ar^{2})}
{12\pi(1+ar^{2})^{2}}+\frac{4B}{3})(\frac{Gr^{3}b}{2(1+ar^{2})}+\frac{4\pi
Gr^{3}p_{r}(r)}
{c^{2}})}{r(c^{2}r-\frac{Gr^{3}b}{1+ar^{2}})}\Bigg]r
\end{equation}

and are enough to obtain the mass-radius relation, the density
profile, and the pressures (tangential and radial). Here,
$b=8\pi\rho_{c}/3$ and $a=1/r_{o}^{2}$. We have checked that the
anisotropy is small (see Figure \ref{pressoes}) in the interval
of central densities. Once the
free parameters $r_{o}$ and $K$ are fitted as functions of the
central density  $\rho_{c}$ the models are completely specified.
We show the resulting mass-radius relation in Figure
\ref{massa-raio}.

\begin{figure}[h!]
 \centering
 \includegraphics[scale=0.625]{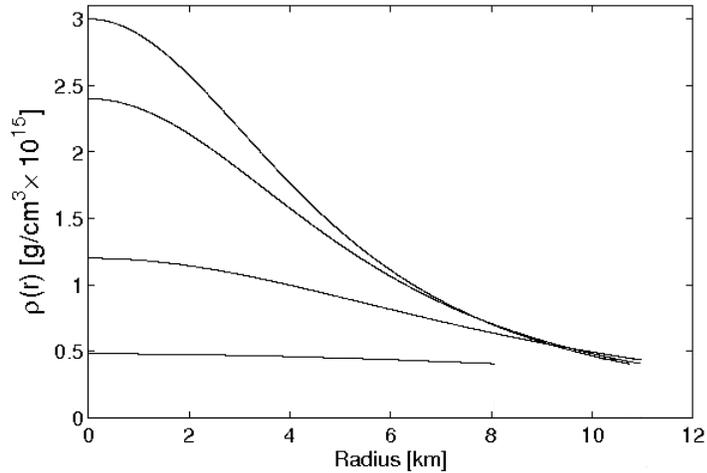}
 \caption{Density profiles of the anisotropic models of Sharma-Maharaj Ref. 16}
 \label{dens_aniso}
\end{figure}

The similarity of the curve in Figure \ref{massa-raio} and the
numerical results by AFO is apparent. However, in spite that
masses and radii are practically the same, anisotropic models are
systematically denser at the center by $25 \%$ or so. For example,
the maximum mass model along the sequence has a central density of
$\simeq 24\times 10^{14}g/cm^{3}$ whereas in the AFO numerical
calculation the value is just $\simeq 20\times 10^{14}g/cm^{3}$.
The density profiles are shown in Figure \ref{dens_aniso}. It is
also apparent the similarities with AFO.

Ruderman\cite{ruderman} argued that the anisotropies could be important for 
densities $\rho >10^{15}g/cm^{3}$. This is consistent with our results.
It should be pointed out, however, that the anisotropy could affect some parameters 
like the maximum mass (the effect is small in our approach) and the redshift, 
as discussed long ago by Bowers and Liang\cite{bowers}.

Only the analytical solution for the anisotropic star is not sufficient for 
a complete description of such an object. This is because it is important to 
explain where the anisotropy comes from in order to produce the differences we 
have shown. Mak and Harko\cite{mak} pointed out that a source of anisotropy 
could be an anisotropic velocity distribution of the particles inside the 
star due, for example, a magnetic field, turbulence or convection.
Perhaps the biggest challenge of the anisotropic model is the stability criterion. 
Chan {\it et al}\cite{chan} showed that, in the onset of instabilities, even small 
anisotropies might drastically change the stability of the system.

We have shown that the pressure anisotropy is small for low central densities 
but becomes larger and larger as the central density increases. Further studies 
are necessary in our approach to verify if all the sequence (e.g. in all central density range) is stable.

\subsubsection{Exact electric field model}

\begin{figure}[h!]
 \centering
 \includegraphics[scale=0.625]{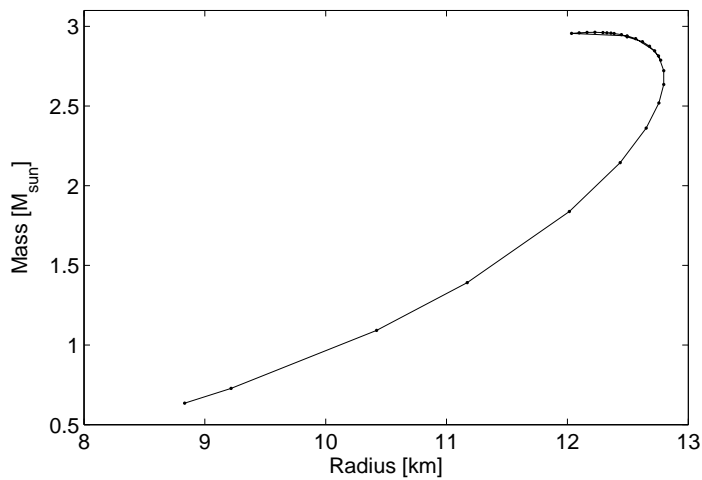}
 \caption{Mass-radius curve for the electric model of Komathiraj-Maharaj Ref. 22}
 \label{massa-raio_kom}
\end{figure}

In the same way as before, we explored the model developed by
Komathiraj and Maharaj\cite{komathiraj} to model strange
stars parametrized by a single
parameter, the central density. This model also has all the
desirable properties cited above. However, the electric field is
an explicit function of the position coordinate, starting from
zero at the center and growing up to the surface. The effect of
this field on the mass-radius relation is to increase the masses
and respective radii. The mass-radius relation is shown in Figure
\ref{massa-raio_kom}.

In this model, we deal with a charged strange star. It is
important to notice that this is quite different from the
assumption made by Usov, since the charge is distributed inside
the whole star. Charged compact stars also have been studied in
many ways by Ray {\it et al.}\cite{malheiro2}. In this work they
have found that a star can have a electric field about $\sim
10^{21}V/m$ in a particular case of polytropic equation of state.
It is interesting to notice that the Komathiraj-Maharaj exact
solution with electric field also provides electric fields at
boundary $ > 1.5\times 10^{20}V/m$.

\section{Conclusions}

We have used a wide set of mathematical solutions (exact and
approximate) in order to model strange quark stars. All these
solutions were parametrized by a single quantity, the central
density. Both simple exact and quasi-exact models were addressed,
with and without extra degrees of freedom (anisotropy and electric
field).

No simple analytical model was found to describe these stars by
keeping the equation of state functionally unchanged (see 2.2.1).
Solutions other than the ones discussed here should be examined
having this description in mind.

\begin{figure}[h!]
 \centering
 \includegraphics[scale=0.625]{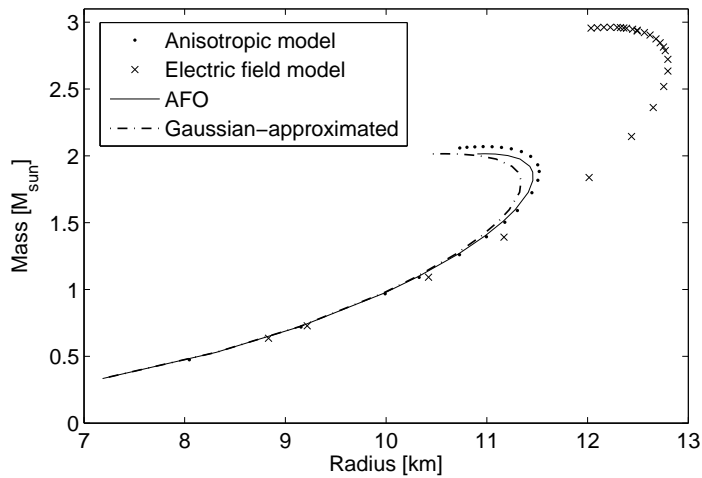}
 \caption{Mass-radius relations for various models.}
 \label{massa-raio-todos-modelos}
\end{figure}

\begin{figure}[h!]
 \centering
 \includegraphics[scale=0.625]{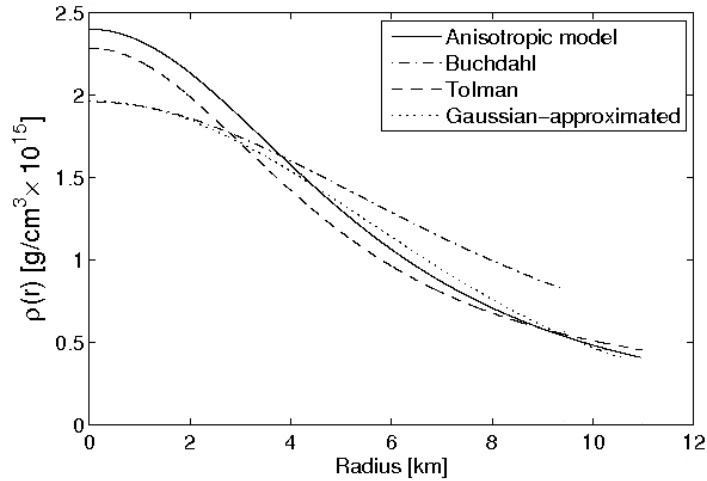}
 \caption{Density profiles of the stars of maximum mass.}
 \label{perfis_dens_massas_maximas}
\end{figure}

In our particular treatment, the quasi-exact solution, we have
used a grid with 20 points, corresponding to 20 central densities
and we have fitted expressions for the free parameters of the
problem that solves analytically the Einstein equations for a
perfect fluid with spherical symmetry and a linear equation of
state. The success of this particular approach can be gauged in
Figure \ref{massa-raio-todos-modelos}, that summarizes the
mass-radius relations of all solutions found and also in Figure
\ref{perfis_dens_massas_maximas} where we show the density
profiles of the stars of maximum mass from some solutions (Tolman,
Buchdahl, gaussian-approximated and anisotropic). We remind that
this work used the simple numerical models of AFO as a benchmark,
but the later has a small number of degrees of freedom, being
based on the MIT bag equation of state. This linear behavior,
however, makes the fluid equation system impossible to be
integrated analytically unless we give an extra equation
overdetermining the system. If this is done, then it is necessary
a match of the overdetermined system.

We have shown that it is possible to find a quasi-exact solution
that is useful to model a strange star in an easy way. We just
give the central density and all other quantities like de mass,
radius, profiles etc. can be found. With these analytical
expressions we can predict many properties of the star defined by
that central density. The errors remain $\leq 1 \%$ in all quantities, 
including the geometry (Fig. 5 and 6).

We can see in the Figure \ref{massa-raio-todos-modelos} that the
gaussian-approximated solution is valid in
the range of low central densities, matching almost exactly to
AFO. The anisotropic model is valid for all
range of central densities and is not very different of AFO. This
is a promising model (see Figure \ref{dens_aniso} and compare it
with Figure \ref{perfil_dens_AFO}; but see also the remarks in the 
section 2.2.4). The electric field model is
also valid, but only in the context of charged quark stars.

This integrability problem does not happen with a system of fluid
equations featuring anisotropic pressure or an electric field.
These new degrees of freedom makes the system integrable. We have
discussed how these results compare to the linear, isotropic,
uncharged models and how they affect the actual stellar features
in practice.

It should be reminded that, in spite of their deep physical
differences, it is too early to dismiss any equation of state for
strange stars. The extraction of the radius from the observation of 
thermal-like emission is still problematic and may contain substantial 
inaccuracies. Taken at face value, however, the determinations have 
rendered small numbers, certainly incompatible with neutron star models 
if true. For example, the objects EXO 0748-676\cite{ozel1}
[$2.10\pm 0.28 M_{\odot},\geq13.8\pm 1.8 km$] and EXO
1745-248\cite{ozel2} [$1.4M_{\odot},11km$] or [$1.7M_{\odot},9km$]
can be described by the anisotropic model (taking into account the
error bars) respectively [$2.0M_{\odot},11.44km$] and
[$1.39M_{\odot},11.0km$] (this last one with a claimed high
precision). We conclude this work pointing out that simple,
economical descriptions of a variety of strange stars can be
constructed and may be useful to explore their properties and
structural behavior in many static and dynamical situations.

\end{document}